\documentclass{elsart}

\usepackage{natbib}

 \usepackage{graphics}

\usepackage{amssymb}

\begin{document}

\begin{frontmatter}



\title{Advances of plasma diagnostics with high-resolution spectroscopy of stellar coronae}


\author{J.-U. Ness}\footnote{Present address: Rudolf Peierls Centre for
Theoretical Physics, University of Oxford, 1 Keble Road, Oxford OX1\,3NP, UK}

\address{Hamburger Sternwarte, Gojenbergsweg 112, 21029 Hamburg, Germany}

\begin{abstract}
X-ray emission from cool stars is an important tracer for stellar activity. The
X-ray luminosity reflects different levels of activity and covers four orders of
magnitude in stars of spectral types M-F. Low spectral resolution provided by
X-ray observations of stellar coronae in the past allowed the determination of
temperature distributions and elemental abundances making use of atomic databases
(listing line emissivities and bremsstrahlung continuum for a given temperature
structure). The new missions XMM-Newton and Chandra carry X-ray gratings providing
sufficient spectral resolution to measure the fluxes of strategic emission lines.
I describe the different approaches applicable to low-resolution and
high-resolution spectra, especially focusing on the new grating spectra with
X-ray lines. From only a few lines it is possible to determine plasma temperatures
and associated densities, to check for any effects from resonant scattering, and to
identify particular abundance anomalies. Line-based temperature- and
density measurements represent only a fraction of the total plasma, but the pressure
environment of different fractions can be probed simply by selection of specific
lines. Selected results are presented covering all aspects of line-based analyses.

\end{abstract}

\begin{keyword}
Stellar atmospheres \sep Stellar activity \sep Main-sequence: late-type stars
\PACS 97.10.Ex \sep 97.10.Jb \sep 97.20.Jg

\end{keyword}

\end{frontmatter}

\section{Introduction}
\label{intro}
\vspace{-.5cm}
 The term activity for the Sun and for late-type stars summarizes phenomena in the
outer atmosphere. Important connections are known between sunspots (places where
magnetic fields pierce the surface and suppress convection) and active regions in
the corona (regions with particularly high temperature and strong X-ray emission).
Also, the appearance of active regions is more frequent with solar maximum and the
X-ray output exhibits the same cycle as the sunspots. The interface between the
surface and the upper corona is the chromosphere, where Ca\,{\sc ii} emission
originates. The intensity of the Ca\,{\sc ii}
emission (measured in the middle of the Ca\,{\sc ii} photospheric absorption line)
sensitively reacts to changes in the solar activity cycle, and is at present (with
few exceptions) the only tracer for stellar activity cycles.\\
 It took severe efforts to actually discover the extreme physical properties of the
solar corona (extremely high temperatures and very low densities). Measurements
of the optical corona revealed that the spectrum is almost identical to the
spectrum of the solar photosphere at all heights, suggesting scattering of
photospheric light by electrons. Since most Fraunhofer lines could not be
identified and the strongest lines were found smeared out, \cite{grotrian31}
concluded that the electrons must have extremely high mean velocities, which are
not consistent with photospheric temperatures. However, only the identification
of emission lines from highly ionized species led to the undoubted conclusion of
the million degree corona \citep[e.g.,][]{edlen,grotrian}. This high temperature
requires X-ray observations in order
to directly look into the million degree plasma of stellar coronae; no contamination
from the stellar photosphere needs be dealt with in this wavelength region. Past
X-ray missions like Einstein and ROSAT were able to discover the ubiquitous
occurrence of hot, tenuous coronae around late-type stars by detection of
considerable X-ray luminosities for {\bf all} late-type stars within the
immediate neighborhood of the Sun \citep[e.g.,][]{schm97}. The X-ray luminosity
is a classical activity indicator and a relation between the X-ray luminosity and
the rotational velocity $v\sin i$ \citep{pal81}
has been established. This suggests magnetic dynamo generation to be involved
in the creation of stellar coronae and therefore also for the solar corona.\\
 Common practice in analyzing X-ray spectra has been based on global fit
approaches (see Sect.~\ref{lowres}). With this method a general trend of plasma
temperatures increasing with activity was found \citep[e.g.,][]{cjm91}. Since
the causes of the heating phenomena have still today not been discovered, this is
an important contribution towards a complete future understanding of the formation and
heating of stellar coronae and the solar corona.\\
 In this spirit, the detailed physical description of stellar coronae is the
next step in order to approach this aim. X-ray spectroscopy of the solar
corona has been applied to measure temperatures and densities in specific
active or quiescent regions. The solar corona was found to be
essentially optically thin and the X-ray spectrum is thus dominated by
emission lines. Important diagnostics tools have been developed, e.g., the
density diagnostics with He-like triplets \citep{gj69}. All these
diagnostics can in principle also be applied to stellar coronae, however,
very sensitive instruments are required in order to provide a decent
S/N at the required spectral resolution. Also, the results have to be interpreted
with the limitation that only average coronal properties can be obtained,
because no spatial resolution is possible. The X-ray missions Chandra and
XMM-Newton provide the ideal instrumental setup with their slitless grating
spectrometers. With these gratings, spectra have been obtained in the last
four years which clearly confirmed that the X-ray spectra of stellar coronae
are also dominated by emission lines originating from highly ionized atomic
transitions. For more active stars continuum emission is found which
is dominated by the bremsstrahlung mechanism \citep[e.g., ][]{ness_alg}. While
the continuum can be
used in order to obtain the plasma temperature of the hottest regions in
the corona, the formation of each individual line reflects the physical
conditions of the plasma regions emitting the respective lines. Since no
individual emission region can be isolated in a stellar corona the line
diagnostics will return averages of all visible emission regions, weighted
with the brightness of each region. This limitation implies that only
typical activity-related physical properties can be
identified. When samples of stellar coronae are investigated, trends between
coronal properties and stellar parameters can be found, uncovering the
underlying physical processes.\\
 This paper will give a review of coronal physical parameters that can be
deduced from spectral line
analysis. While in the past X-ray spectra did not have the power to measure
individual lines, the new gratings aboard Chandra and XMM-Newton allow individual
line fluxes to be measured for the first time. This requires new analytical
approaches. In principle, a complete model spectrum can be synthesized from tables
containing all our knowledge of the atomic physics (atomic databases) and be compared
with measured spectra of any spectral resolution. This method (global fitting) will
be limited by the quality of the spectrum (when applied to low-resolution spectra)
or by the quality of the atomic database in use (when applied to high-resolution spectra).
I will first describe how low-resolution spectra have been analyzed
and what could be learnt and then address a number of aspects which have been
deduced from the analysis of individual lines.
\vspace{-.5cm}
\section{Analysis of low-resolution spectra}
\label{lowres}
\vspace{-.5cm}
 While the earliest X-ray missions allowed only the detection of the X-ray intensity
in a broad energy band, more refined missions had some spectral resolution based on the
energy-sensitivity of the detectors (CCDs, proportional counters). Although
individual spectral features could not directly be seen a lot of useful information
could still be obtained by convolving model spectra to the instrumental spectral
resolution. These model spectra basically contain the relevant atomic physics and within
a surprisingly good range the physical parameters could be optimized in order to find
good agreement with the measured spectra.\\
 In order to construct a model spectrum an atomic database is needed, which contains
the information on the formation of lines induced by atomic transitions under the
assumption of an optically thin plasma (i.e., only the production of photons is
described, but no absorption; see also Sect.~\ref{opt}). Since the coronal plasma is
in principle dominated by collisional ionizations and excitations, it is sufficient
for the modelling of stellar coronal spectra to assume the 'coronal' approximation.
Going beyond this assumption would need more refined efforts.\\
 The parameters put into a model are temperatures, emission measures, and elemental
abundances. The temperature is the main parameter entering a spectral model. It will
affect the ionization fraction and (along with the density) the population of the
excited levels. Also the contribution of a continuum, which consists of bremsstrahlung,
recombination, and two-photon continuum, depends on the temperature. Since
nature does not provide isothermal plasma, a temperature distribution has to be assumed.
This can be approximated by using, e.g., three isothermal components, which each carry a
weighting factor in terms of an emission measure value (specifying the amount of
emitting material in the corona at the given temperature). Three spectral models are
then constructed and will be co-added for the final model. The number of temperature
components can be chosen arbitrarily high, but including additional temperature
components makes only sense when an improvement in agreement with the measurements
can be accomplished. In contrast to a (generally low) number of isothermal components,
a smooth temperature distribution can be assumed and optimized \citep[e.g., ][]{schm90}.
The next physical parameter is a set of elemental abundances. All lines originating
from ions of the same element are linearly scaled by the value of its abundance.
In the models the elemental abundances and the temperature components are modelled
simultaneously, but a sufficient number of lines must lie in the spectral region for
sensible constraints on the abundance. Note, however, that changes in the elemental
abundances will also affect the bremsstrahlung continuum, because in hot plasma,
highly ionized metals will insert a considerable number of additional electrons.
Within the ranges of abundances now typically found in stellar coronae the
bremsstrahlung continuum changes only by a few percent for typical hot coronal
temperatures.\\
 With these spectral models it was possible to establish a temperature-activity relation
\citep{cjm91,schm90}. Also, it was found that the hotter average temperatures
in more active stars are mainly caused by an additional hotter temperature component,
while only a slight shift of the cooler temperature component was noticed
\citep{guedel97}. This can be explained by an increasing number of active regions with
increasing activity as in the solar activity cycle.\\
 The methods applied to model low-resolution spectra can in principle also be applied
to high-resolution spectra (which show individual emission lines). However,
the accuracy of the available atomic databases then imposes the major limitations,
while the same method applied to low-resolution spectra was limited by the quality
of the spectra. The improvement of the quality of the results has thus been pushed
to the limits of the databases and further progress can only be made by improving
the atomic data.\\
 An alternative approach is to measure the line fluxes of strategically chosen
individual lines and compute line fluxes reflecting specific physical aspects.
In the next section I will discuss some aspects which can be addressed by measurement
of line fluxes and line flux ratios.
\vspace{-.5cm}
\section{Analysis of high-resolution spectra}
\label{highres}
\vspace{-.5cm}
 A high-resolution spectrum in the present context is defined as a spectrum which allows
one to resolve a minimum number (at least five to ten) of individual emission lines.
Taking the present X-ray missions
this implies that the CCD spectra (ACIS-S, ACIS-I, EPIC-PN, and EPIC-MOS) are considered
low-resolution spectra, while the grating spectra (LETGS, HETGS, and RGS) are considered
high-resolution spectra.\\
The same methods applied to low-resolution spectra, especially the global fit approaches
can just as well be applied to high-resolution spectra and the results gain accuracy
from the improved spectral resolution, but not beyond the quality of the atomic databases.
At the moment, the limitations of the results are determined by the quality
of the databases alone. However, the quality of global fits can be improved further by
excluding wavelength regions where there are high degrees of uncertainties in the 
databases by calculating the fit goodness parameter as demonstrated by \cite{aud03}.
With this approach one can avoid to a large extent misidentifications of measured line
features belonging to lines not listed in the databases. In the latter case a global fit
would seek physical conditions pulling up other line fluxes listed at nearby wavelengths,
while line-based approaches would leave these features as unidentified, ignoring them
for further interpretation.\\
The three X-ray gratings cover the wavelength ranges 1--40\,\AA\ (the LETGS goes up
to 175\,\AA) with different spectral resolution and sensitivity. The strategic lines
in this wavelength region are the H-like and He-like lines of Si, Mg, Ne, O, N,
and C (altogether 24 lines). Also, a number of Fe L-shell and K-shell lines are
measurable with these instruments. It is possible to obtain a large amount of
information from these few lines without the use of global models, which use thousands
of lines simultaneously.\\
 For the interpretation of line fluxes and line ratios the same atomic databases must be
used. The line-based analysis uses the strongest lines with the smallest
uncertainties (constrained by both theoretical calculations and laboratory measurements),
but these strong lines might also be blended with fainter lines from complicated ions,
e.g., lines of Fe. The blending can be significant, e.g., for the Ne He-like lines
\citep{nebr} and in these cases the limitations are essentially the same as in global
models. Accounting for the blending lines is intrinsically implemented in the global
fit approach, while a line-based approach has to carefully predict the blending lines.
\vspace{-.5cm}
\subsection{Measurement of opacities}
\label{opt}
\vspace{-.5cm}
Before any analyses based on the information obtained from the atomic data\-bases can be
carried out, one has to assure that any measured photon rate actually represents the
photon production rates. In principle, photons produced in lower layers might be absorbed
in higher layers and re-emitted into other directions (scattering). The solar corona is
commonly assumed to be optically thin, however, the strongest resonance lines with high
radiative excitation probabilities might place considerable absorption cross sections
into the line of sight. Absorbed photons will be re-emitted, but not necessarily back
into the line of sight and some photons will be re-emitted back to the stellar surface
and are thus effectively lost. In all cases, with no balance of scattering out of the line
of sight and into the line of sight (called ''effectively optically thin''), these
''resonant scattering'' effects would significantly distort the measured line fluxes
compared to those produced. This distortion can be modelled, but the modelling
requires assumptions about the structure of the absorbing layers (e.g., spherical
geometries) but coronal plasma can have extreme geometries (especially when active
regions are involved), so the modelling would become extremely
complicated. It is therefore common practice to neglect resonant scattering effects,
but this can be tested. In principle one could easily see resonant scattering effects
when comparing measured (affected) resonance lines with measured forbidden lines, which
can be considered to be 100\% thin. The ratio of a resonance line and a forbidden
line can be compared with a corresponding theoretical ratio from the databases
predicting optically thin fluxes.
In order to eliminate any temperature- and abundance effects, the choice of
lines should focus on lines of the same element and the same ionization stage. A
prominent example is the ratio of two Fe\,{\sc xvii} lines at 15\,\AA\ and 15.27\,\AA.
The latter is a forbidden line (oscillator strength $f=0.6$ in contrast to
$f=2.6$ for the 15\,\AA\ line) and the ratio $\lambda$15.27/$\lambda$15 will increase
with increasing opacity effects. \cite{ness_opt} have compared a large number of
stellar $\lambda$15.27/$\lambda$15 (and other) ratios with each other and found
the measured ratios all higher than theoretical predictions for an optical thin plasma
(suggestive of significant opacity effects). However, they found no systematic trend
with activity yielding similar ratios for all stars in their sample. This means that
the amount of emitting plasma has no effect on the line ratios, and so \cite{ness_opt}
concluded that the higher measured ratios rather imply erroneous theoretical ratios than
identical optical depths for all kinds of stellar
coronae. \cite{testa_opt} analyzed the ratio Ly$_\beta$/Ly$_\alpha$ for the ions of
oxygen and neon for a sample of stars, but found only two exceptions from the
zero-optical depth scenario. However, their claims of unique first-time findings of
resonant scattering effects in stellar coronae are confirmed by both ions only for one
corona (IM\,Peg).\\
The general conclusion is that it appears to be reasonable to neglect resonant scattering
effects in coronal plasma, but one has to check individual spectra.
\vspace{-.5cm}
\subsection{Abundance anomalies}
\label{aabun}
\begin{figure}[ht]
\resizebox{\hsize}{!}{\includegraphics{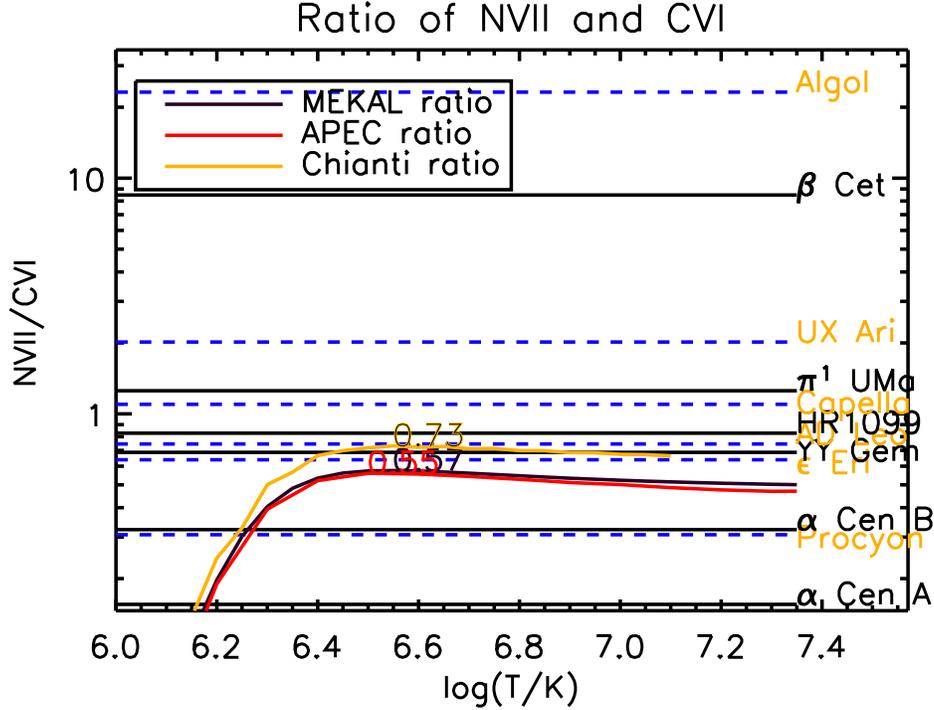}}
\caption{\label{c6_n7}Predicted ratio of N\,{\sc vii}/C\,{\sc vi} at solar abundances as
a function of temperature. Clearly, some extremely high measured ratios require increased
nitrogen abundances.}
\end{figure}

\vspace{-.5cm}
For coronae resembling the solar corona an abundance anomaly called The FIP effect was
discovered with EUVE, Chandra, and XMM-Newton. This effect has long been known
from the solar corona and all elements with a first ionization
potential (FIP) lower than 10\,eV ($\sim 1216$\,\AA, which is the wavelength of the
hydrogen Ly$_\alpha$ line) are overabundant compared to photospheric abundances. The
heating or transport mechanisms obviously prefer to deal with species, which are
ionized (possibly by photoionization from Ly$_\alpha$ line photons) which can then
couple to magnetic fields. For many more active stars an inverse FIP effect has been
found with XMM-Newton \citep[e.g.,][]{aud03}, which is puzzling. The detailed background
to the FIP and inverse FIP effects are complicated, but a recent approach by
\cite{laming04} explains the fractionation quite naturally by ponderomotive forces
arising as upward propagating Alfv{\'e}n waves from the chromosphere transmit or reflect
upon reaching the chromosphere-corona boundary.\\
 While about half of these results have been determined by the use of global fits to
high-resolution spectra, some methods have been developed to obtain elemental abundances
using the specific advantages of measuring individual line fluxes \citep[see summary
in][]{guedelaarev}. \cite{abun} developed
a method using the H-like to He-like line ratios (see Sect.~\ref{temp}) to construct the
temperature distribution independently of elemental abundances. All remaining
discrepancies
from a spectrum constructed from this temperature distribution then reflect abundance
effects. \cite{telli04} compared global methods applied to limited spectral ranges
containing bright lines \citep{aud03} with a line-based approach, and found good
agreement between these methods.\\
An interesting effect was discovered by \cite{cno}. The lack of any carbon
line in the Chandra LETGS spectrum of Algol raised suspicion of an abundance effect,
since the nitrogen lines were well detected. Fig.~\ref{c6_n7} demonstrates that
N\,{\sc vii}/C\,{\sc vi} flux ratios greater than one cannot be explained by a
temperature effect, but must be explained by abundance anomalies.
This effect is well in line with expectations from stellar evolution. Since Algol is
evolved, the anomalous abundance pattern reflects dredged-up CNO cycled material. All
stars for which an enhanced N\,{\sc vii}/C\,{\sc vi} ratio was detected are evolved stars,
while the other stars show normal abundances, with $\alpha$\,Cen and Procyon showing low
ratios, probably due to low coronal temperatures.
\vspace{-.5cm}
\subsection{Plasma temperatures}
\label{temp}
\begin{figure}[ht] 
\resizebox{\hsize}{!}{\includegraphics{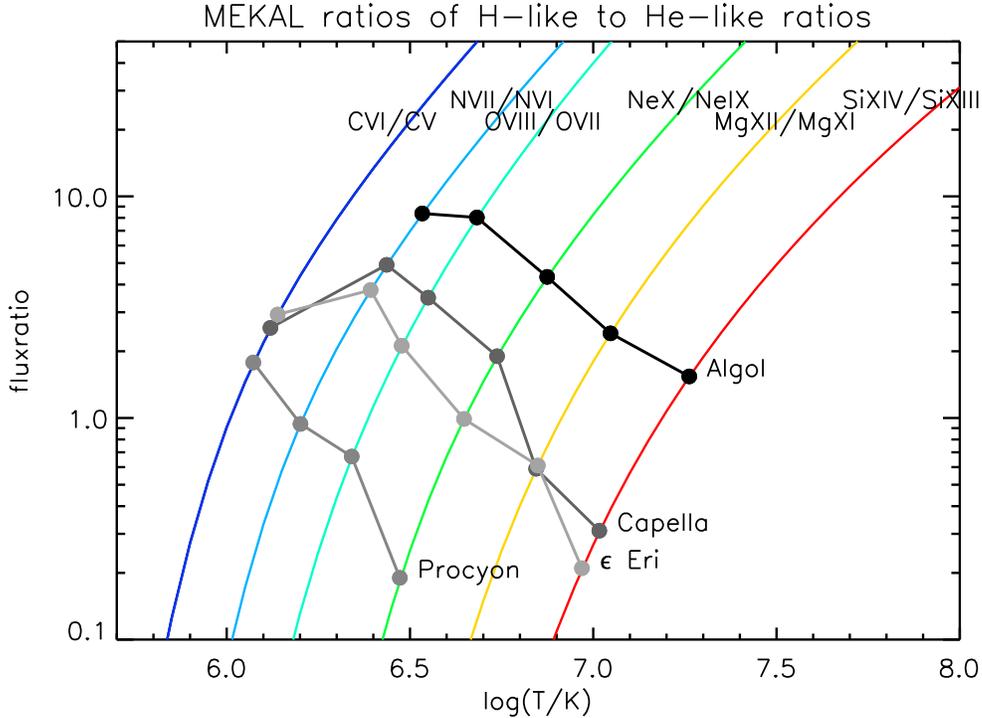}}
\caption{\label{lyrats}Temperature-sensitive line ratios based on H-like and He-like line fluxes.}
\end{figure}

\vspace{-.5cm}
 While the global fits allow one to obtain immediately a complete temperature
distribution, individual lines can probe temperatures for specific regions of the
temperature distribution. Again, a smart choice of line ratios allows one to eliminate
other effects
besides temperatures, e.g., elemental abundances. From the available strong lines the
ratio of the H-like to one of the He-like triplet (r, i, f) lines (usually the
resonance line r or the sum of all three lines) of the same element is
temperature-sensitive due to the ionization balance. In hotter plasma the H-like line
will be stronger while in cooler plasma the He-like lines dominate. The ratios can then
be compared to theoretical predictions and probe all plasma emitting the respective
lines. In Fig.~\ref{lyrats} I show the theoretical predictions of H-like to He-like
line ratios for different elements \citep[see][]{abun}. A steep increase of the ratios
can be recognized
indicating a very sensitive temperature diagnostic. In addition I include measured
line flux ratios for different stars, and it can be seen that the stars selected have
quite different temperature distributions. While Algol has systematically higher
line ratios (indicative of higher temperatures), Procyon has only a high carbon H-like
to He-like line ratio, while the Si lines are produced at higher temperatures than
found in Procyon's corona. These ratios provide a good starting point for constructing
temperature distributions, where the ratios can be used as interpolation points.
The shape of the emission measure distribution has to reproduce the measured
line ratios, and no abundance effects interfere, because all line ratios are independent
of the elemental abundances.
\vspace{-.5cm}
\subsection{Densities}
\label{dens}
\vspace{-.5cm}
 Density measurements are not possible with low-resolution spectra, because the effects
from densities are too subtle to significantly affect a low-resolution spectrum.
Therefore, no density analyses have been carried out for stellar X-ray coronae and
structural information was only available from eclipsing systems, where one component
is X-ray dark \citep[e.g., for $\alpha$\,CrB or Algol][estimated densities from the
spatial distribution of intensity]{guedel03,algolflare}. From measurements of the
emission measure EM, the total emitting volume $V$ can be inferred if densities $n_e$ are
known by simply applying the relation EM$=0.85n_e^2V$, which defines the volume emission
measure; here, a homogeneous geometry is assumed. Densities are also needed in order to
apply loop scaling laws (developed for the Sun) in order to investigate whether stellar
coronae can be considered as scaled-up versions of solar active regions or whether new
concepts have to be developed. Again, a geometry has to be assumed, e.g., a set of
identical loop-like structures.\\
The measurement of densities from high-resolution spectra exploits the increasing number
of collisions with increasing electron densities. In a low-density plasma transitions
with low de-excitation probabilities (forbidden lines) will still show up. With increasing
electron densities the upper levels of these transitions are increasingly subject to
collisionally induced further excitations into higher levels with higher radiative
de-excitation probabilities and those lines will show up instead. All density analyses
are based on either measuring the appearance of the latter lines \citep[e.g.,
Fe\,{\sc xxi} lines in EUVE spectra:][]{mason} or the disappearance of the former; or
both \citep[He-like triplets:][]{gj69}.\\
The density measurements are carried out from line flux ratios in order to eliminate abundance
and temperature effects. A number of Fe\,{\sc xxi} lines which are expected to show up in
high-density plasma and an Fe\,{\sc xxi} resonance line (at 128.73\,\AA) were measured
with EUVE. The ratios of the fluxes in the former lines with those in the latter were
analysed by \cite{dupr93}, who reported extremely high densities for Capella, while, e.g.,
\cite{schmitt94} found no evidence at all for deviations from the low-density limits
for Procyon. The difficulty with these diagnostics has been described by \cite{ness_dens}.
One can never say
whether an emission feature at the expected wavelength of a density-sensitive line
actually corresponds exactly to this line, or whether it comes from (an) unidentified
line(s). \cite{ness_dens} investigated
several Fe\,{\sc xxi} line ratios for several stellar coronae using the LETGS (Chandra)
and found not a single star with consistently high densities from all line ratios. Some
ratios suggested higher densities (when believing the measured line fluxes to belong to
the expected lines), but they were ruled out again by other line ratios, measured at the
same time from the same ion (therefore formed in exactly the same environments).\\
 Analyses of the He-like triplets measure the ratio of a forbidden line, f
($^3$S$_1$--$^1$S$_0$), versus an intercombination line, i ($^3$P$_1$--$^1$S$_0$),
\citep[f/i][]{gj69,ness_dens}, where the f line will put its photons into the i line
with increasing densities. The principle is the same for all He-like ions from different
species, formed in different plasma regions with different temperatures. The Chandra and
XMM-Newton gratings can measure the He-like triplet lines from Si\,{\sc xiii} ($Z=14$)
down to C\,{\sc v} ($Z=6$) and plasma regions with temperatures ranging from 15\,MK
down to 1\,MK can be probed, measuring densities in the range
$10^9$--$10^{14}$\,cm$^{-3}$. Those He-like triplets formed at high temperatures
probe only high densities (above $10^{12}$\,cm$^{-3}$), while low-temperature ions
measure only lower densities ($\sim 10^{10}$\,cm$^{-3}$); this leaves two cases
unexplored: low densities in hot plasma and high densities in cool plasma. From
the O\,{\sc vii} measurements of stellar coronae the case of high densities at low
temperature can be excluded, but the case of low densities at high temperatures
remains unexplored.
\vspace{-.5cm}
\section{Conclusions}
\vspace{-.5cm}
 The analysis of emission line fluxes from grating X-ray spectra is a powerful tool
complementing
global fit approaches. It is possible to survey the temperatures in different regions
of the temperature distribution, identify abundance anomalies, recognize effects from
resonant scattering, and measure densities. Some of these issues can only be addressed
with emission line measurements, especially the densities. \cite{ness_dens} measured
O\,{\sc vii} and Ne\,{\sc ix} densities and \cite{testa04} measured the Mg\,{\sc xi}
densities, and the combined results suggest that all three ions originate from
different pressure regions. Since the coronal structures implied (generally believed to be
loop-like arches) usually do not extend higher than the pressure scale heights of the
individual stars, each loop must have constant pressure,
and the different pressures from the different density diagnostics thus imply that
different classes of loops exist. The O\,{\sc vii} loops are characterized by
low pressures and low temperatures (thus small scale heights), and these loops are
found to occur
in stellar coronae in all stages of activity. In contrast to this, the Ne\,{\sc ix}
and the Mg\,{\sc xi} loops have higher pressures (higher temperatures {\bf and} higher
densities) and occur with increasing number in more active stars
(characterized by higher X-ray surface fluxes). It appears reasonable to conclude
that a standard cool temperature corona always exists, while active regions containing
hotter plasma are
a privilege of the more active stars.

\vspace{-.5cm}
\section*{Acknowledgments}
\vspace{-.7cm}
I thank Prof. Carole Jordan for stimulating discussion about the paper.
I acknowledge financial support from Deutsches Zentrum f\"ur Luft- und
Raumfahrt e.V. (DLR) under 50OR98010.
The comments by the first referee, Dr. Manuel Guedel, are highly appreciated
and improved the quality of the paper.
\vspace{-.5cm}

\bibliographystyle{elsart-harv}
\bibliography{hist,jhmm,astron,jn}

\end{document}